\documentclass[prd,superscriptaddress,10pt,twocolumn,preprintnumbers]{revtex4}

\usepackage[latin1]{inputenc}
\usepackage{graphicx}
\usepackage[english]{babel}

\usepackage{amsmath}
\usepackage{amssymb}
\usepackage{amsfonts}
\usepackage{colordvi}
\usepackage{psfrag}
\usepackage{color}

\begin{document}
\def\cL{{\cal L}}
\def\be{\begin{equation}}
\def\ee{\end{equation}}
\def\bea{\begin{eqnarray}}
\def\eea{\end{eqnarray}}
\def\beq{\begin{eqnarray}}
\def\eeq{\end{eqnarray}}
\def\tr{{\rm tr}\, }
\def\nn{\nonumber \\}
\def\e{{\rm e}}

%
%


\def\bef{\begin{figure}}
\def\eef{\end{figure}}
\newcommand{\ans}{ansatz }
\newcommand{\eeqn}{\end{eqnarray}}
\newcommand{\bd}{\begin{displaymath}}
\newcommand{\ed}{\end{displaymath}}
\newcommand{\mat}[4]{\left(\begin{array}{cc}{#1}&{#2}\\{#3}&{#4}
\end{array}\right)}
\newcommand{\matr}[9]{\left(\begin{array}{ccc}{#1}&{#2}&{#3}\\
{#4}&{#5}&{#6}\\{#7}&{#8}&{#9}\end{array}\right)}
\newcommand{\matrr}[6]{\left(\begin{array}{cc}{#1}&{#2}\\
{#3}&{#4}\\{#5}&{#6}\end{array}\right)}
\newcommand{\cvb}[3]{#1^{#2}_{#3}}
\def\lsim{\raise0.3ex\hbox{$\;<$\kern-0.75em\raise-1.1ex
e\hbox{$\sim\;$}}}
\def\gsim{\raise0.3ex\hbox{$\;>$\kern-0.75em\raise-1.1ex
\hbox{$\sim\;$}}}
\def\abs#1{\left| #1\right|}
\def\simlt{\mathrel{\lower2.5pt\vbox{\lineskip=0pt\baselineskip=0pt
           \hbox{$<$}\hbox{$\sim$}}}}
\def\simgt{\mathrel{\lower2.5pt\vbox{\lineskip=0pt\baselineskip=0pt
           \hbox{$>$}\hbox{$\sim$}}}}
\def\unity{{\hbox{1\kern-.8mm l}}}
\newcommand{\eps}{\varepsilon}
\def\ep{\epsilon}
\def\ga{\gamma}
\def\Ga{\Gamma}
\def\om{\omega}
\def\omp{{\omega^\prime}}
\def\Om{\Omega}
\def\la{\lambda}
\def\La{\Lambda}
\def\al{\alpha}
\newcommand{\ov}{\overline}
\renewcommand{\to}{\rightarrow}
\renewcommand{\vec}[1]{\mathbf{#1}}
\newcommand{\vect}[1]{\mbox{\boldmath$#1$}}
\def\tm{{\widetilde{m}}}
\def\mcirc{{\stackrel{o}{m}}}
\newcommand{\Dm}{\Delta m}
\newcommand{\dm}{\varepsilon}
\newcommand{\tanb}{\tan\beta}
\newcommand{\nbar}{\tilde{n}}
\newcommand\PM[1]{\begin{pmatrix}#1\end{pmatrix}}
\newcommand{\up}{\uparrow}
\newcommand{\down}{\downarrow}
\def\omE{\omega_{\rm Ter}}
%

\newcommand{\Dsusy}{{susy \hspace{-9.4pt} \slash}\;}
\newcommand{\DCP}{{CP \hspace{-7.4pt} \slash}\;}
\newcommand{\mc}{\mathcal}
\newcommand{\gr}{\mathbf}
\renewcommand{\to}{\rightarrow}
\newcommand{\gtc}{\mathfrak}
\newcommand{\wh}{\widehat}
\newcommand{\br}{\langle}
\newcommand{\kt}{\rangle}


\def\lsim{\mathrel{\mathop  {\hbox{\lower0.5ex\hbox{$\sim$}
\kern-0.8em\lower-0.7ex\hbox{$<$}}}}}
\def\gsim{\mathrel{\mathop  {\hbox{\lower0.5ex\hbox{$\sim$}
\kern-0.8em\lower-0.7ex\hbox{$>$}}}}}

\def\nn{\\  \nonumber}
\def\de{\partial}
\def\brf{{\mathbf f}}
\def\bbf{\bar{\bf f}}
\def\bF{{\bf F}}
\def\bbF{\bar{\bf F}}
\def\bA{{\mathbf A}}
\def\bB{{\mathbf B}}
\def\bG{{\mathbf G}}
\def\bI{{\mathbf I}}
\def\bM{{\mathbf M}}
\def\bY{{\mathbf Y}}
\def\bX{{\mathbf X}}
\def\bS{{\mathbf S}}
\def\bb{{\mathbf b}}
\def\bh{{\mathbf h}}
\def\bg{{\mathbf g}}
\def\bla{{\mathbf \la}}
\def\bmu{\mathbf m }
\def\by{{\mathbf y}}
\def\bmu{\mbox{\boldmath $\mu$} }
\def\bsig{\mbox{\boldmath $\sigma$} }
\def\bunity{{\mathbf 1}}
\def\cA{{\cal A}}
\def\cB{{\cal B}}
\def\cC{{\cal C}}
\def\cD{{\cal D}}
\def\cF{{\cal F}}
\def\cG{{\cal G}}
\def\cH{{\cal H}}
\def\cI{{\cal I}}
\def\cL{{\cal L}}
\def\cN{{\cal N}}
\def\cM{{\cal M}}
\def\cO{{\cal O}}
\def\cR{{\cal R}}
\def\cS{{\cal S}}
\def\cT{{\cal T}}
\def\eV{{\rm eV}}

\title{Quantum Ekpyrotic mechanism in Fermi-bounce curvaton cosmology}


\author{Andrea Addazi}

\author{Antonino Marciano}

\affiliation{ Center for Field Theory and Particle Physics \& Department of Physics, Fudan University, 200433 Shanghai, China}

\date{\today}
\begin{abstract}
\noindent 
Within the context of the Fermi-bounce curvaton mechanism, we analyze the one-loop radiative corrections to the four fermion interaction, generated by the non-dynamical torsion field in the Einstein-Cartan-Holst-Sciama-Kibble theory. We show that contributions that arise from the one-loop radiative corrections modify the energy-momentum tensor, {\it mimicking} an effective Ekpyrotic fluid contribution. For these reasons, we call this effect {\it quantum Ekpyrotic} mechanism. This leads to the dynamical washing out of anisotropic contributions to the energy-momentum tensor, without introducing any new extra Ekpyrotic fluid. We discuss the stability of the bouncing mechanism and derive the renormalization group flow of the dimensional coupling constant $\xi$, checking that any change of its sign takes place towards the bounce. This enforces the theoretical motivations in favor of the torsion curvaton bounce cosmology as an alternative candidate to the inflation paradigm.  
\end{abstract}


\maketitle

\section{Introduction}
\noindent 
The cosmological bounce scenario, as originated by standard model matter fields, is a viable alternative to inflation not yet experimentally ruled out. The idea of the bounce is that the cosmological dynamics emerges from a pre-Big Bang universe, which solves the cosmological singularity. Almost scale-invariant perturbations are generated during the contracting cosmological phase, thus satisfying CMB observations \cite{Planck}. A possible origin of the bounce can be recovered within the context of the Einstein-Cartan-Holst-Sciama-Kibble theory (ECHSK) \cite{Torsion,Torsion2,Torsion3,Perez:2005pm}. In the first order formalism, one must allow for a torsionful part of the spin-connection, since gravity is coupled to fermion fields. In this framework, the torsion field does entail propagating degrees of freedom, and can be integrated-out as an auxiliary field. By virtue of the torsion-fermion coupling, new effective four-fermion interactions arise. A sort of effective Nambu--Jona-Lasinio (NJL) model can emerge, but from a very different dynamical origin than the one addressed in the QCD case. Depending on the sign of the fermion-torsion coupling, this can provide a new attractive or repulsive contribution to the energy conditions. In the repulsive phase, with a certain choice of the torsion-fermion coupling, such a term can contribute to the energy-momentum tensor, triggering a bounce at a related critical energy density scale. Possible bounce cosmology scenarios were investigated in Refs.~\cite{Bambi:2014uua,Alexander:2014eva,Lucat:2015rla,Addazi:2016rnz}.

A typical issue for bounce scenarios in cosmology is the generation of large $O(1)$ anisotropic terms in the energy-momentum tensor, which is incompatible with the cosmological isotropy of the CMB.  A way out to this problem requires the introduction of a new exotic fluid, the Ekpyrotic fluid, provided with an energy-density steeply scaling with the Universe scale factor, {\it i.e.} $\rho_{\rm Ek}= \rho_{0}a^{-n}$ with $n>6$. The Ekpyrotic fluid does dominate during the Bounce critical scale, suppressing anisotropic contributions to the background dynamics. However, both the classical Bounce and the Ekpyrotic mechanism are based on a classical analysis. During the Bounce stage, quantum corrections are expected to be large and potentially they may completely change and wash-out the Bounce dynamics, as well as the Ekpyrotic classical solution to the anisotropies' problem. 

This highly motivates the analysis of quantum corrections to the Ekpyrotic mechanism in some specific Bounce cosmology models. In this paper, we will analyze the one-loop quantum corrections to the four-fermion term generated by torsion, in a Friedmann-Lema\^itre-Robertson-Walker (FLRW) background. We will show that quantum corrections will induce extra corrections to the classical bare energy-momentum, previously unconsidered. We find that these new radiative terms have theoretically two healthy consequences: $i)$ they favor the classical Bounce, generating contributions that violate the Null Energy Condition (NEC); $ii)$ they generate new terms mimicking the effect of the Ekpyrotic fluid. This result can likely mean that the four-fermion curvaton mechanism can induce the Bounce without introducing any new exotic matter field. This hypothesis seems to be theoretically appealing, since it avoids adding to the theoretical framework any particle that cannot be accommodated in the Standard Model of particle physics, or in any Grand Unified extensions of it. 

\section{The theory}
\noindent
We follow the theoretical framework and the conventions introduced in Refs.~\cite{Torsion, Torsion2, Torsion3, Perez:2005pm, Bambi:2014uua, Lucat:2015rla, Addazi:2016rnz, Alexander:2014eva, Alexander:2014uaa, Freidel:2005sn, Alexander:2008vt, Pop, BD2, Magueijo:2012ug, BD, M, ABC}. This amounts to consider a generalization of the Einstein-Hilbert action with a topological term, the Holst action for gravity in the Palatini formalism, which allows to couple gravity to chiral fermions. The theory can be coupled to a Dirac field $\psi$, and to the related field $\overline{\psi}=\left(\psi^{*}\right)^{T}\gamma^{0}$. The Dirac action involves the Dirac matrices, $\gamma^I$ with $I=0,\ldots,3$ and $\gamma^5$. The action for pure gravity can be cast in terms of the gravitational field $g_{\mu \nu}=e_{\mu}^I e_{\nu}^J \eta_{IJ} $, where $e^I_\mu$ is the tetrad/frame field (with inverse $e_I^\mu$ and determinant $e$), and the Lorentz connection $\omega^{IJ}_\mu$. The curvature of $\omega^{IJ}_\mu$, namely 
$$F^{I J}_{\mu\nu}=2 \partial_{[\mu}\omega^{IJ}_{\,\nu]} + \left[ \omega_\mu, \omega_\nu \right]^{IJ}\,,$$ 
is the triadic projection of the Riemann tensor.

\begin{figure}[t]
\centerline{ \includegraphics [width=0.9 \columnwidth]{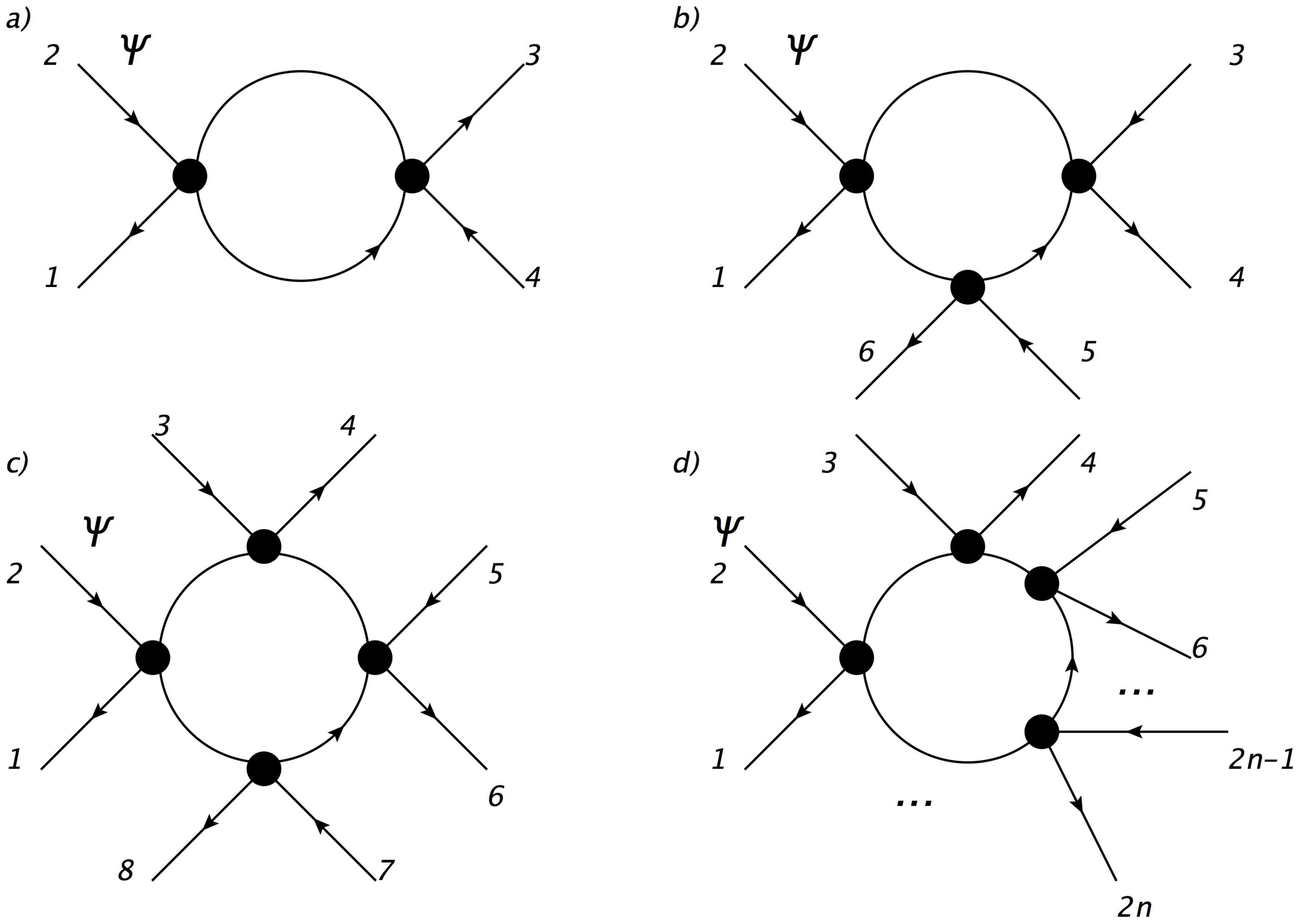}}
\vspace*{-1ex}
\caption{a) One-loop correction to the four fermions self-interaction term; b) one loop correction to the six-points self-interaction term; c) one loop correction to the eight-points self-interaction term; c)  one loop correction to the 2n-points self-interaction term}
\label{plot}   
\end{figure}

\begin{figure}[t]
\centerline{ \includegraphics [width=0.9 \columnwidth]{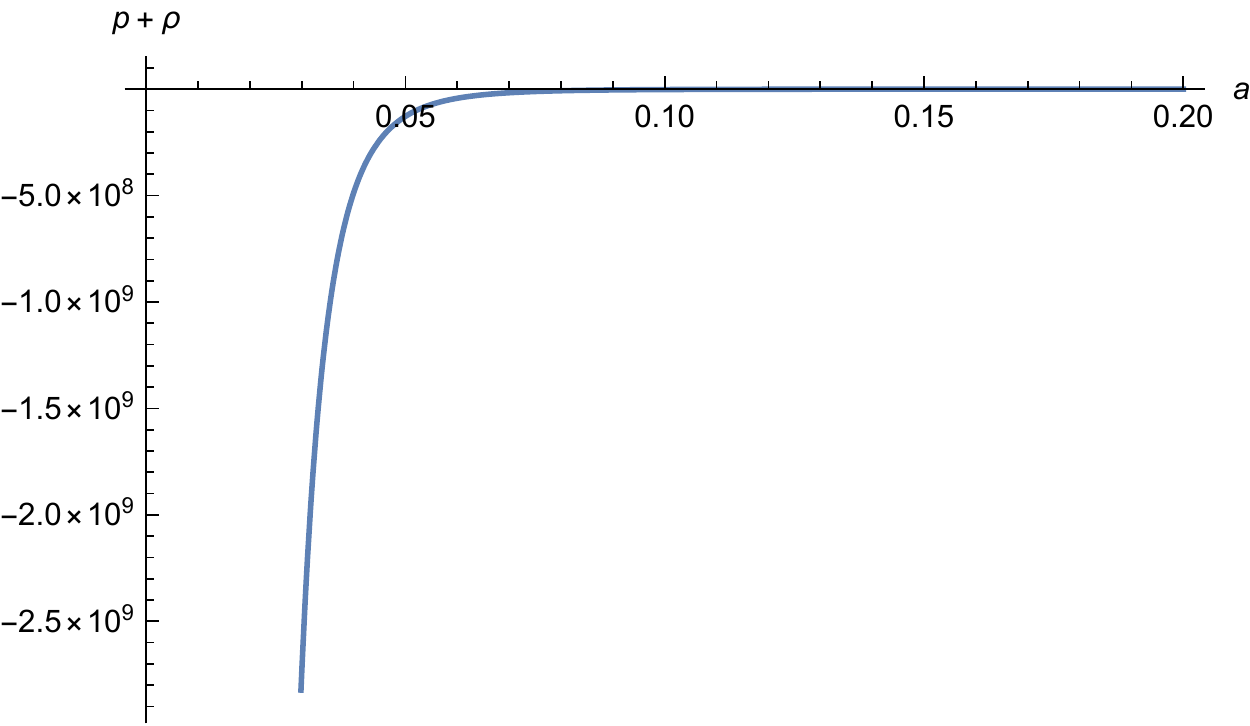}}
\vspace*{-1ex}
\caption{The $p+\rho$ is displayed in Planckian units, as a function of the scale factor $a$ and for the choice of the parameters $\mu=\xi=\kappa=1$ and $q_{5}=2$. 
At the bounce, the NEC $p+\rho \geq 0$ is violated.  }
\label{plot}   
\end{figure}

The total action involves the Einstein-Cartan-Holst (ECH) action plus the non-minimal covariant Dirac action. Notice that, in absence of the gravitational Holst topological term, the whole theory provided with torsion and minimally-coupled fermions is referred to in the literature as the Einstein-Cartan-Sciama-Kibble theory (see {\it e.g.} Refs.~\cite{Torsion, Torsion2,Torsion3}). In the Palatini first order formalism, the ECH action casts (see {\it e.g.}~\cite{Alexander:2014eva}), 
\begin{eqnarray}
\label{nonminimalaction1}
&&{S}_{\rm Holst} =
\frac{1}{2 \kappa} \int_{M}\!\!  d^{4}x \;|e| \, e^{\mu}_{I}e^{\nu}_{J}
P^{IJ}_{\ \ \ KL}F^{\ \ KL}_{\mu \nu}(\omega)\, 
\end{eqnarray}
In Eq.~\eqref{nonminimalaction1}, $\kappa = 8 \pi G_{\rm N}$ is the square of the reduced Planck length, $\epsilon_{IJKL}$ denotes the Levi-Civita symbol, while the tensor in the internal indices $P^{IJ}_{\ \ \ KL}=\delta^{[I}_{K} \delta^{J]}_{L} - \epsilon^{IJ}_{\ \ KL}/ (2 \gamma)$ involves the Barbero--Immirzi parameter $\gamma$ and is invertible for $\gamma^2\neq -1$. The Dirac action for massless fermions reads $S_{\rm Dirac} = \!\frac{1}{2 }  \int d^{4}x |e| \mathcal{L}_{\rm Dirac}$, with
\begin{eqnarray}
\label{nonminimalaction2}
\mathcal{L}_{\rm Dirac} = \!\frac{\imath}{2 } \overline{\psi}\!\left(1-\frac{\imath}{\alpha}\gamma_{5}\right)\!\gamma^{I}e^{\mu}_{I} \nabla_{\mu}\psi + {\rm h.c.}
\, .
\end{eqnarray}
In Eq.~\eqref{nonminimalaction2} $\alpha$ denotes a real parameter called non-minimal coupling. For $\alpha\!=\!\gamma$ we may recover the Einstein-Cartan action, namely $S_{\rm ECH}\!\!=\!\!S_{\rm GR} \!+\! S_{\rm Dirac}$. An additional term is also present, which reduces to the Nieh-Yan invariant (see {\it e.g.} Ref.~\cite{M}) when the second Cartan structure equation is satisfied. Within the framework of the Holst action in Eq.~(\ref{nonminimalaction1}), the minimal coupling  can be recovered in the limit $\alpha \rightarrow \pm\infty$. Phenomenologically allowed values of the parameters $\alpha$ and $\gamma$ are recovered from the four fermion axial-current Lagrangian (\ref{interact}), from measurements of lepton-quark contact interactions \cite{Freidel:2005sn, deBlas:2013qqa}. 

The presence of fermions induces a torsional part of the connection to enter the non-minimal ECH action. Torsion is non-dynamical in this theory, and can be integrated out once the second Cartan structure equation is solved. This requires to introduce the contortion tensor $C_\mu^{IJ}$, which is defined by $(\nabla_\mu-\widetilde{\nabla}_\mu ) V_I = C_{\mu \,I}^{\ \ J}\, V_J\,,$ with $\widetilde{\nabla}_\mu$ denoting the covariant derivative compatible with the tetrad $e^I_\mu$ and $V_J$ a vector in the internal space. The Cartan equation expresses the contortion tensor $C_\mu^{IJ}$ in terms of the fermions' currents and the tetrad, namely 
\begin{eqnarray} \label{carta}
&& e^\mu_I \, C_{\mu JK} = \frac{\kappa}{4} \, \frac{\gamma}{\gamma^2 +1} \, \left( \beta \, \epsilon _{IJKL}\ J^L - 2 \theta \, \eta_{I[J} \, J_{K]}  \right) \,, \nonumber \\
&&J^L=\overline{\psi} \gamma^L \gamma_5 \psi 
\, .
\end{eqnarray}
The coefficients appearing in Eq.~\eqref{carta} are related to the free parameters of the non-minimal ECH theory, $\beta=\gamma+1/\alpha$ and $\theta=1-\gamma/\alpha$. Deploying the solution in (\ref{carta}), the non-minimal ECH action recasts in terms only of the metric compatible variables, further including a novel interaction term that captures the new physics: 
\begin{eqnarray}
\label{ECH}
S_{\rm ECH}= S_{\rm GR} + S_{\rm Dirac}
+S_{\rm Int}\,.
\end{eqnarray}
In Eq.~\eqref{ECH}, the Einstein-Hilbert action involves the mixed-indices Riemann tensor $R_{\mu\nu}^{IJ}=F_{\mu\nu}^{IJ}[\widetilde{\omega}(e)]$, namely 
\begin{eqnarray}
S_{GR}= \frac{1}{2 \kappa} \int_{M} \!\!\! d^4 x |e| e^\mu_I e^\nu_J R_{\mu\nu}^{IJ} \,, 
\end{eqnarray}
the Dirac action $S_{\rm Dirac}$ on curved space-time recasts in terms of the metric compatible variables 
\begin{eqnarray}
S_{\rm Dirac}= \frac{\imath}{2} \int_{M} \!\!\! d^4 x |e| \ 
\overline{\psi} \gamma^I e^\mu_I  \widetilde{\nabla}_\mu \psi +{\rm h.c.}\,,
\end{eqnarray}
while the interacting term reads 
\begin{eqnarray} \label{interact}
S_{\rm Int} \!=\! -\xi \kappa\! \int_{M} \!\!\! d^4 x |e| \,J^L\, J^M\, \eta_{LM}\,,
\end{eqnarray}
with the coefficient $\xi$ being now a function of the fundamental parameters of the theory, namely 
\begin{eqnarray}
\label{xi}
\xi:= \frac{3}{16} \!\frac{\gamma^2}{\gamma^2+1}\! \left(1 + \frac{2}{\alpha \gamma} -  \frac{1}{\alpha^2} \right)\,.
\end{eqnarray}
In few words, net effect of having coupled fermions to gravity in the first order formalism is to have recovered, in the metric-compatible formalism, an extra four fermion term. That means that the structure of General Relativity remains untouched, but that the matter part of the Einstein equations acquires at tree-level a self-interaction term, the coupling constant of which, $\xi$, can be either positive or negative. 

For the sake of completeness, we show in this section the energy-momentum tensor components, {\it i.e.}
\begin{eqnarray} \label{tenfe}
\hspace{-0.25cm}
T^{\rm fer}_{\mu\nu}\!=\! \frac{1}{4}  \overline{\psi} \gamma_I e^I_{( \mu} \imath \widetilde{\nabla}_{\nu )} \psi +{\rm h.c.}  -
 g_{\mu\nu} \mathcal{L}_{\rm fer}\,.
\end{eqnarray} 
%

\section{One-loop corrections}
\noindent 
Since non-dynamical torsion provides an interaction term, we may reconstruct at one-loop the quantum effective action of the matter part of the theory. For simplicity, in order to calculate the these terms, we assume the background to be FLRW, which in conformal coordinates reads 
\begin{equation}
\label{dsd}
ds^{2}=a^{2}(-d\eta^{2}+d{\bf x}\cdot d{\bf x})\,,
\end{equation}
with $a=a(\eta)$ scale factor and $\eta$ conformal time.

The four-fermion vertex is 
\begin{equation} 
\label{fourvertex}
\mathcal{V}_{4}=-\imath \, {\xi} \kappa\, \gamma_{\mu}\gamma_{5} \otimes \gamma^{\mu}\gamma_{5}
\end{equation}

The Feynman propagator in position space between the points $x$ and $\tilde{x}$ is denoted as $\imath S_{F}(x,\tilde{x})$, and reads
\begin{equation}
\label{ypp}
\imath S_{F}(x,\tilde{x})= (a\tilde{a})^{-\frac{D-1}{2}} \frac{\Gamma(D/2-1)}{4 \pi^{D/2}}\, \imath \gamma^a \partial_a\, \frac{1}{\Delta x_{++}^{D-2}(x,\tilde{x})}\,,
\end{equation}
having denoted $\tilde{a}=a(\tilde{\eta})$, and
\begin{equation}
\label{Dx}
\Delta x_{++}^{D-2}(x,\tilde{x})=( \left\Vert \vec{x}-\vec{\tilde{x}}\right\Vert ^{2}-\left|\eta-\tilde{\eta}\right|^{2})^{D-2}\,,
\end{equation}
in which the two spacetime points are labelled with $x=\{\eta, \vec{x} \}$ and $\tilde{x}=\{\tilde{\eta}, \tilde{\vec{x}} \}$. 
Eq.~\eqref{ypp} is an straightforward result to be calculated, easier than the estimate of the massless scalar propagator, because of the conformal nature of massless fermions in any dimension --- see {\it e.g.} \cite{Janssen}. Assuming fermion matter content to be massless around the bounce is a viable approximations, given that the critical energy density at bounce is much above the condensate phase of the Higgs. We disregard in this preliminary study to include Yukawa couplings of fermionic species to the Higgs, leaving a refined analysis to a forthcoming analysis \cite{SNM}.

The sign of the coupling constant $\xi$, expressed in Eq.~\eqref{xi} in terms of the bare parameters of the ECHSK theory, $\alpha$ and $\gamma$, individuates two phases at the cosmological level. For $\xi>0$, the theory is in a Fermi liquid phase, with a tree-level repulsive interaction that has been studied as a source of dark energy \cite{ABC}, or to drive inflation \cite{Addazi:2017qus}. In stead, the super-conductive phase corresponds to the branch of values $\xi<0$, for which a bouncing scenario has been taken into account  \cite{AP,Alexander:2008vt,Pop,Alexander:2014eva,Alexander:2014uaa,Addazi:2016rnz}. Therefore, we will proceed taking into account negative values of $\xi$ in what follows. In Sec.~\ref{stab}, we will come back to the issue of the quantum stability of the bounce, and analyze the running of the $\xi$ parameter around the bounce.

\subsection*{Lowest n-point contributions}
\noindent 
We start considering the 2-point and the 3-point 1-loop contributions to the effective action, derived from the evaluation of the diagrams in Fig.1-a and Fig.1-b. 
For the former one, derived from the Wick contraction of two pairs of axial currents, one has to calculate the tensorial contribution $\Pi^{\mu \nu}_{2}$, which is defined 
\begin{eqnarray}
\Pi_{2}^{\mu \nu}= -{\rm Tr}_{x'} [  \gamma^\mu \gamma^5 S_{F}(x,\tilde{x}) \gamma^\nu \gamma^5  S_{F}(\tilde{x},x)]\,,
\end{eqnarray}
where we have the tensorial structure is the one specified in \eqref{fourvertex}, the fermion propagator $S_{F}(x,\tilde{x})$ is the one defined in \eqref{ypp}, ${\rm Tr}_{x'}$ denotes the trace over the internal spinorial indices, and integration over spacetime points $x'$. The massless propagator in \eqref{ypp} can be expanded according to
$S_{F}(x,\tilde{x})=\gamma^a \,S_a(x,\tilde{x})$, which allows to factorize the tensorial structure from the integral, {\it i.e.}
\begin{eqnarray}
\Pi_{2}^{m r}= \!\!\!\!&& - 4 (\eta^{a m} \eta^{b r} - \eta^{r m} \eta^{a b} + \eta^{b m} \eta^{ a r } ) \times  \nonumber\\ 
&&\, (-\imath \xi \kappa)^2 \int \sqrt{-g}\,  d^4 x \, S_a(x,\tilde{x}) S_b(\tilde{x},x)\,. 
\end{eqnarray}
This amounts to calculate two types of contribution:
\begin{eqnarray}
\bar{\Pi}_{2}^{ab}= - (\xi \kappa)^2\, \int \sqrt{-g}\,  d^4 x \,\, S^a(x,\tilde{x}) S^b(\tilde{x},x)\,, \label{pi0}\\
\bar{\Pi}_{2}=  - (\xi \kappa)^2\, \int \sqrt{-g}\,  d^4 x \,\, S_a(x,\tilde{x}) S^a(\tilde{x},x)\,.\label{pi2}
\end{eqnarray}
For the former, in generic $D$ dimensions, we find 
\begin{eqnarray}
\bar{\Pi}_{2}^{ab}= \!\!\!\!\!&&  - (\xi \kappa)^2  \, a^{-(D-1)} 
\int \, d^D \, \tilde{x}\, \frac{[\Gamma(D/2-1)]^2}{16 \pi^D}\, \times \nonumber\\
&& \left( \imath  \partial^a\, \frac{1}{\Delta x_{++}^{D-2}(x,\tilde{x})} \right) \, 
\left( \imath  \tilde{\partial}^b\, \frac{1}{\Delta x_{++}^{D-2}(\tilde{x},x)} \right)
\nonumber\\
&&= - (\xi \kappa)^2  \, a^{-(D-1)} 
\int \, d^D \, \tilde{x}\, \frac{[\Gamma(D/2-1)]^2}{16 \pi^D}\, \times \nonumber\\
&& \, \frac{1}{\Delta x_{++}^{D-2}(\tilde{x},x)} \, 
\partial^a \tilde{\partial}^b\, \frac{1}{\Delta x_{++}^{D-2}(x,\tilde{x})}  \,.  \nonumber
\end{eqnarray}
For $D=4$, the above formulas yield the expressions
\begin{eqnarray}
  \bar{\Pi}_{2}^{ab} = \!\!\!\!\!&& - (\xi \kappa)^2  \, a^{-3} 
 \int \, d^4 \, \tilde{x}\, \frac{1}{16 \pi^4}\, \frac{1}{\Delta x_{++}^{2}(\tilde{x},x)} \,  \times \nonumber\\
&& \, 
\partial^a \tilde{\partial}^b\, \frac{1}{\Delta x_{++}^{2}(x,\tilde{x})}  =   \frac{(\xi \kappa)^2} { 2 \pi^4 a^{3} 
} \int \!\! d^4 \, {\zeta}\, \frac{\zeta^a \, \zeta^b}{{\zeta}^8}  \nonumber \\
&& = - \frac{ (\xi \kappa \,\mu^2)^2 
}{4\pi^4 a^{3} 
}\!\! \int \!\! d^4\Omega \, n^a\, n^b   
\!= \! - \frac{ (\xi \kappa\, \mu^2)^2 
}{16\pi^2 a^{3} 
} \eta^{ab}\!,
\end{eqnarray}
%
$\mu$ representing the sliding scale, and having used the fact that $\int \!\! d^4\Omega \, n^a\, n^b= \frac{\pi^2}{2} \,\eta^{ab}$. In the renormalization group (RG) flow of the coupling constant $\xi$, we shall retain this dependence, according to the hierarchy $\mu^2\leq 1/(\xi\kappa)$. \\

For the quantity in \eqref{pi2}, the expression recasts as
\begin{eqnarray}
\bar{\Pi}_{2}&&\!\!\!\!\!\!\! =  - (\xi \kappa)^2  \, a^{-(D-1)}\! 
\int \! d^D \, \tilde{x}\, \frac{[\Gamma(D/2-1)]^2}{16 \pi^D}\, \times \nonumber\\
&&\!\!\!\!\!\!\! \left( \imath \partial_a\, \frac{1}{\Delta x_{++}^{D-2}(x,\tilde{x})} \right) \, 
\left( \imath  \tilde{\partial}^a\, \frac{1}{\Delta x_{++}^{D-2}(\tilde{x},x)} \right)
\nonumber\\
&&\!\!\!\!\!\!\! = - (\xi \kappa)^2  \, a^{-(D-1)}\! 
\int \! d^D \, \tilde{x}\, \frac{[\Gamma(D/2-1)]^2}{16 \pi^D}\, \times \nonumber\\
&& \!\!\!\!\!\!\!  \frac{1}{\Delta x_{++}^{D-2}(\tilde{x},x)} \, 
\partial_a \tilde{\partial}^a\, \frac{1}{\Delta x_{++}^{D-2}(x,\tilde{x})}  \,.  \nonumber
\end{eqnarray}
For $D=4$, making use of 
\begin{equation}
\partial^2 \, \frac{1}{\Delta x_{++}^{2}} = 4 \imath \pi^2
\, \delta^4(x-\tilde{x})\,,
\end{equation}
we finally obtain
\begin{eqnarray}
\!\!\!\bar{\Pi}_{2} 
 = \frac{\imath (\xi \kappa)^2 }{4 \,\pi^3 \, a^{3}
 } \! \int \! d^4 \, \tilde{x} \, \frac{1}{\Delta x_{++}^{2}(\tilde{x},x)} \, 
\delta^4(x-\tilde{x})=0 \,.
\end{eqnarray}

It is trivial to realize that 3-point contribution vanishes, because of the properties of the gamma matrices, {\it i.e.} 
\begin{eqnarray}
&&\Pi_{3}^{\mu \nu \rho}= \nonumber\\
&&-{\rm Tr}_{\tilde{x}, x'} [  \gamma^\mu \gamma^5 S_{F}(x,\tilde{x}) \gamma^\nu \gamma^5  S_{F}(\tilde{x},x')\gamma^\nu \gamma^5  S_{F}(x',x)]=0\,. \nonumber
\end{eqnarray}
The four-point function is recovered from the expression
\begin{eqnarray}
&&\Pi_{4}^{\mu_1 \mu_2 \mu_3 \mu_4}= -{\rm Tr}_{x_2, x_3,x_4} [  \gamma^{\mu_1} \gamma^5 S_{F}(x_1,x_2)  \times \nonumber \\ 
&& \gamma^{\mu_2} \gamma^5 S_{F}(x_2,x_3)  \gamma^{\mu_3} \gamma^5 S_{F}(x_3,x_4)   \gamma^{\mu_4} \gamma^5 S_{F}(x_4,x_1)]\,. \nonumber
\end{eqnarray}
%
%
The one-loop corrections arising from the 2-point and 4-point functions, as depicted in Fig.~1-a and Fig.~Fig.~1-c (and accounting for the symmetry degeneration factor of the internal lines), leads to the form of the effective potential 
\begin{eqnarray}
\label{veff}
V_{\rm eff}= && \!\!\!\!\! V_0 +\xi\kappa\, (\overline{\psi} \gamma_5 \gamma_a \psi)^2 \left( 1+  \frac{\xi \kappa\mu^2 }{\pi^2 a^{3}} \right) +\nonumber\\
&& \!\!\!\!\!\!\!  
 \frac{(\xi\kappa)^4\ln(\xi \kappa\mu)}{8 \pi^2\, a^3}  
[(\overline{\psi} \gamma_5 \gamma_a \psi)^2]^2 
 \!+\!O\left( (\xi \kappa\mu^2)^6 \right).
\end{eqnarray}
The constant $V_0$ represents a shift of the energy density, required to avoid negative values, which would be inconsistent with the Einstein equations. This is a short way to ensure the theoretical consistency of this approach. Differently, in a fully general approach that takes into account also the inclusion of hypercharge fields, positive energy densities would be provided by the energy density of radiation. It is also worth to notice that the vacuum polarization diagram in Fig.1-b cannot contribute to the energy density, because of the tensorial structure of the vertex in Eq.~\eqref{fourvertex}. \\ 

From the conservation of the chiral current, which is implied by the assumption of having massless fermions, it follows that in \eqref{veff} each axial current redshifts as the volume factor. Isotropy the background requires the spatial component to be vanishing. This can be checked for slowly varying backgrounds, by expressing the axial current in terms of the scalar and pseudoscalar bilinears, and the vector current, {\it i.e.} by using the Pauli-Fierz identity
\begin{eqnarray}
&& (\overline{\psi} \gamma_5 \gamma^a \psi)  (\overline{\psi} \gamma_5 \gamma_a  \psi) = \nonumber\\
&& \qquad  \qquad = (\overline{\psi}  \psi)^2 - (\overline{\psi} \gamma_5 \psi)^2 + (\overline{\psi} \gamma^I  \psi)  (\overline{\psi}  \gamma_I  \psi)  \, . \nonumber
\end{eqnarray} 
The vector current can be then checked to posses only non-vanishing temporal components --- see {\it e.g.} Ref.~\cite{Alexander:2011hz} --- by contracting the fermion Feynman propagator with the Dirac matrices, and extracting the dimensional regularized part of the coincident limit. Therefore, we can recast the quadratic terms entering the effective potential by means of the expression  
\begin{eqnarray}
(\overline{\psi} \gamma_5 \gamma_a \psi)^2 =({\psi}^\dagger \gamma_5 \psi)^2=\frac{Q_5^2}{a^6}\,, \quad {\rm with} \quad Q_5\sim q_5\mu^3\,.
\end{eqnarray} 

These arguments immediately lead to the conclusion that an Ekpyrotic quantum effect is generated at one-loop, considering the 2-point contribution to the effective action. The bounce is not spoiled by this effect. With a natural choice of the conformal factor normalized at the bounce, the 2-point contribution to the one-loop correction retains a numerical factor subdominant with respect to tree-level contribution. At the same time, around the bounce the extra term (to the energy potential) washes out anisotropies, thanks to the dependence 
\begin{eqnarray} 
\rho^{(2)}_{\rm Ek}= \frac{\!\!\rho_{0}}{ a^{9}}\qquad {\rm with} \qquad \rho_0= \frac{( Q_5\,\xi \kappa\mu)^2}{256 \pi^4} \,.
\end{eqnarray} 
The $4$-point amplitude then contributes to the effective potential with a dumping factor that redshifts as $a^{15}$, and thus enhances the Ekpyrotic behavior. We can further check that generic $n$-point contributions do not spoil this effect. 

\subsection*{$2n$-point contributions}
\noindent 
Calculations can be carried out for the $2n$-point contributions to the effective one-loop action (see Fig.1-d). One should indeed consider all the possible amplitudes arising from 
\begin{eqnarray}
\!\!\!&&\!\!\!\!\!\!\Pi_{n}^{\mu_1 \dots \mu_{2n}}= \nonumber \\
&& \!\!\!\!\!\!=-{\rm Tr}_{x_1,...x_{2n-1}} [  \gamma^{\mu_1} \gamma^5 S_{F}(x,x_1) \dots \gamma^{\mu_{2n}} \gamma^5  S_{F}(x_{2n-1},x)] \,.\nonumber \\
\end{eqnarray}
For large $n$, these terms behave as 
\begin{eqnarray}
&&\!\!\!\!\!\!\Pi_{n}^{\mu_1 \dots \mu_{2n}}={\rm (tensorial \ structure)} \times \frac{(\xi \kappa)^{2n}}{a^3\, 2 \pi^2\, (2\mu)^{2n} (2n-1)!  }\,,  \nonumber
\end{eqnarray} \\ 
thus entailing the asymptotic contribution to the potential at $2n$-order
\begin{eqnarray}
V_{\rm eff}^{(2n)} \sim \frac{(2 \xi \kappa)^{2n}}{a^3 \, 2 \pi^2 (2\mu)^{2n} }\, \left[ (\overline{\psi} \gamma_5 \gamma^a \psi)^2 \,\right]^{n}  \,. \nonumber
\end{eqnarray} 
Summing for the potential, we find 
$$V_{\rm eff}\,\sim\, V_0 +$$
$$\!\xi\kappa (\overline{\psi} \gamma_5 \gamma^a \psi)^2\!+\! \sum_n (-1)^{n} \frac{ (\xi \kappa)^{2n}}{a^3 \, 2 \pi^2 (2\mu)^{2(n-2)}}\, \left[ (\overline{\psi} \gamma_5 \gamma^a \psi)^2 \,\right]^{n}$$
\begin{equation}\label{nnan} \sim V_0\!+\! \xi\kappa (\overline{\psi} \gamma_5 \gamma^a \psi)^2 \!+\! \frac{\mu^4}{a^3 \, 2 \pi^2[1+\frac{(\xi \kappa)^2}{ 2\mu^{2}} (\overline{\psi} \gamma_5 \gamma^a \psi)^2 ]^2} \,. 
\end{equation}
The potential can be finally recast in order to explicit the scale factor dependence:
\begin{eqnarray}
V_{\rm eff} \sim &&\!\!\! V_0+ \xi \kappa  \frac{Q_5^2}{a^6} + \frac{\mu^4}{a^3 \, 2 \pi^2[1+\frac{(\xi \kappa)^2}{ 2 \mu^{2}} \frac{Q_5^2}{a^6} ]^2}  \nonumber\\
\sim && \!\!\! V_0+  \frac{ \xi\kappa \mu^6}{a^6} + \frac{\mu^4}{a^3 \, 2 \pi^2[1+\frac{(\xi \kappa \mu^2)^2}{2 a^6} q_5^2 ]^2} \,. \nonumber
\end{eqnarray} 
It follows that the bounce cannot be spoiled in this scenario, but that actually extra contributions arise that enhance the quantum Ekpyrotic mechanism.

\section{RG flow and stability}
\label{stab}
\noindent 
Consistency requirements with CMB observables impose that the bounce must take place at critical energies not below the GUT scale. This set-up the values of the critical energy density to be $\rho_{c}\simeq (\xi\kappa)^{-2}\simeq M_{\rm GUT}^4$. Correspondently, we might wonder that the sliding scales could approach the value $\mu\simeq M_{GUT}$ at the bounce, and eventually induce a change of sign of the the renormalized value of $\xi$. Our estimate shows that this is not the case, since the (predominant in the one-loop calculation) $2$-point correction, namely 
\begin{eqnarray}
\xi_{\rm ren.}=\xi \left( 1+  \frac{\xi \kappa\mu^2}{\pi^2 
} + O(\xi \kappa\mu^2)^2 \right)\,,
\end{eqnarray}
maintains the sign at lowest approximation. The $2$-point contribution yields indeed only an order $O(1/10)$ numerical suppression factor at the bounce. This conclusion can be strengthened looking at the calculation of the $2n$-point amplitude contribution. In general, the numerical factor driving RG flow of $\xi$ can be summing  the coupling constant dependence of the vertex-operators that enter the $2n$-point function, and the coefficients of the fermion Feynman propagators. From the calculations we have already shown while deriving the effective potential, it is then immediate to convince ourselves that the higher order coefficients of $\xi_{\rm ren.}$ will actually decrease, as soon as more interaction vertices are accounted for in the one-loop contribution to the effective action, and that the sum can be summed, without encountering a Landau pole at the bounce. 

\section{Conclusions}
\noindent 
We have shown that a {\it quantum} Ekpyrotic scenario emerges from the calculation of the one-loop quantum effective action. The lowest-order vertex interaction is already enough to produce this remarkable effect. This result has been achieved assuming chiral symmetry unbroken at the energy density scale at which the bounce takes place, a scenario accomplished in several unification theories, and naturally envisaged in the non-condensate phase of the Higgs. An important step in order to test this result would be to reproduce this analysis involving anisotropic metrics. The technical limitation in dealing with quantum field theory on Bianchi backgrounds forbids at the present to pursue this goal. 

\section*{Appendix}
\noindent
The results we obtained on the stability of the bounce and the occurrence of a quantum ekpyrotic scenario hinge on some straightforward but lengthly calculations, the details of which are reported in this section. 
\subsection{The $4$-function at $1$-loop}
\noindent
To recover the four-point function $\Pi_{4}^{\mu_1 \mu_2 \mu_3 \mu_4}$ we shall calculate the trace over eight Dirac matrices. Projecting on the Clifford algebra indices provides
\begin{eqnarray}
& \Pi_{4}^{m_1\,m_3\, m_5\, m_7 }= - 4 \, [ \eta^{m_1 m_2} \eta^{m_3 m_4} \eta^{m_5 m_6} \eta^{m_7 m_8} \nonumber \\
&-\eta^{m_1 m_3} \eta^{m_2 m_4} \eta^{m_5 m_6} \eta^{m_7 m_8} + \eta^{m_1 m_3} \eta^{m_2 m_5} \eta^{m_4 m_6} \eta^{m_7 m_8} \nonumber \\
&- \eta^{m_1 m_3} \eta^{m_2 m_5} \eta^{m_4 m_7} \eta^{m_6 m_8} + \eta^{m_8 m_3} \eta^{m_2 m_5} \eta^{m_4 m_7} \eta^{m_6 m_1}\nonumber \\
 &- \eta^{m_8 m_2} \eta^{m_3 m_5} \eta^{m_4 m_7} \eta^{m_6 m_1} + \eta^{m_8 m_3} \eta^{m_2 m_4} \eta^{m_5 m_7} \eta^{m_6 m_1}\nonumber \\
 &- \eta^{m_8 m_3} \eta^{m_2 m_5} \eta^{m_4 m_7} \eta^{m_6 m_1}
 +  \eta^{m_8 m_3} \eta^{m_2 m_5} \eta^{m_4 m_6} \eta^{m_7 m_1}\nonumber \\
  &-\eta^{m_3 m_2} \eta^{m_8 m_5} \eta^{m_4 m_6} \eta^{m_7 m_1} +
  \eta^{m_3 m_2} \eta^{m_5 m_4} \eta^{m_8 m_6} \eta^{m_7 m_1}
 \nonumber \\
 & - \eta^{m_3 m_2} \eta^{m_5 m_4} \eta^{m_6 m_7} \eta^{m_8 m_1} 
  + \eta^{m_8 m_2} \eta^{m_5 m_4} \eta^{m_6 m_7} \eta^{m_1 m_3}  \nonumber \\
 & -  \eta^{m_8 m_2} \eta^{m_4 m_6} \eta^{m_5 m_7} \eta^{m_1 m_3}  +
 \eta^{m_8 m_2} \eta^{m_4 m_6} \eta^{m_5 m_1} \eta^{m_7 m_3}  ] \nonumber \\
& \!\!\!\!\!\!\! \times  (-\imath \xi \kappa)^4 \int \sqrt{-g}\,  d^4 x_2 \, \int \sqrt{-g}\,  d^4 x_3 \, \int \sqrt{-g}\,  d^4 x_4  \nonumber \\
&\times\,  S_{m_2}(x_1, x_2) \,  S_{m_4}(x_2, x_3) \,  S_{m_6}(x_3, x_4) \,  S_{m_8}(x_4, x_1) \, . 
\end{eqnarray} 
The multiple integral term reshuffles as 
\begin{eqnarray}
& I_{m_2\, m_4\, m_6\, m_8}:= \prod \limits_{h=1}^{3} \int \sqrt{-g}\,  d^4 x_{h+1} \prod \limits_{s=1}^{4}  S_{m_{2s}}(x_{s}, x_{|s+1|_4}) \nonumber \\
 & =\frac{1}{32 \pi^8\, a^3 } \prod \limits_{h=1}^{3} \int  d^4 x_{h+1} \prod \limits_{s=1}^{4}  \partial_{m_{2s}}^{(s)} \frac{1}{\Delta_{++}^2(x_{s}, x_{|s+1|_4})}\,,
\end{eqnarray}
where $|\dots|_p$ stands for modulo $p$ and $\partial^{(s)}$ for the derivative with respect to the $m_{2s}$ component of the $x_s$ coordinate. We further recast the integral through a series of passages:
\begin{eqnarray}
&\!\!\! \prod \limits_{h=1}^{3} \int  d^4 x_{h+1} \prod \limits_{s=1}^{4}  \partial_{m_{2s}}^{(s)} \frac{1}{\Delta_{++}^2(x_{s}, x_{|s+1|_4})} \!=\! \int  d^4 x_{2}  \, \partial_{m_2}^{(1)} \frac{1}{(x_{1}- x_{2})^2} \nonumber\\
& \times  \int  d^4 x_{3}  \, \partial_{m_4}^{(2)} \frac{1}{(x_{2}- x_{3})^2}\, \int  d^4 x_{4}  \, \partial_{m_6}^{(3)} \frac{1}{(x_{3}- x_{4})^2}\, \partial_{m_8}^{(4)} \frac{1}{(x_{4}- x_{1})^2} = \nonumber\\
&- 2^4 \, \int  d^4 x_{2}  \,  \frac{(x_1-x_2)_{m_2}}{(x_{1}- x_{2})^4} \, \int  d^4 x_{3}  \,  \frac{(x_2-x_3)_{m_4}}{(x_{2}- x_{3})^4}\, \int  d^4 x_{4}  \,  \frac{(x_4-x_3)_{m_6}}{(x_{4}- x_{3})^4}  \nonumber\\
& \times \frac{(x_4-x_1)_{m_8}}{(x_{4}- x_{1})^4}= 8 \pi^2 \eta_{m_6 m_8} \!\int \! d^4 x_{3}   \int \! d^4 x_{2}   \frac{(x_2-x_1)_{m_2}}{(x_{2}- x_{1})^4} \,  \frac{(x_2-x_3)_{m_4}}{(x_{2}- x_{3})^4} \nonumber\\
& \times \int  \frac{d |y|}{|y-(x_{3}- x_{1})|^3}= 4 \pi^4 \eta_{m_2 m_4} \eta_{m_6 m_8}\, \int d^4z \, \left( \int  \frac{d |y|}{|y-z|^3} \right)^2 =\nonumber\\
& 2 \pi^6 \, \eta_{m_2 m_4}  \eta_{m_6 m_8}\,\int  \frac{d|z|}{|z|}= \pi^6 \, \eta_{m_2 m_4}  \eta_{m_6 m_8}\, \ln\left( \frac{1}{\xi \kappa \mu^2} \right) \nonumber \,,
\end{eqnarray}
where in the last hand-side we have used as ultraviolet regulator $(\xi \kappa)^{-\frac{1}{2}}$. Combining all the terms together, we find the contribution to the effective potential at $4$-point order. 

\subsection{The $2n$-function at $1$-loop}
\noindent
The $2n$-point function $\Pi_{n}^{\mu_1 \dot \mu_{2n}}$ requires the knowledge of the trace over $4n$ gamma matrices. In general, one can show that 
\begin{eqnarray}
{\rm Tr}[\gamma^{m_1}\dots \gamma^{m_{4n}}]= 4 \, [ (-1)^{|P|} \, \eta^{\sigma(m_1) \sigma(m_2)} \dots \eta^{\sigma(m_{2n-1}) \sigma(m_{2n})} ]\,, \nonumber 
\end{eqnarray}
where $\sigma$ denotes all the possible non-cyclic permutations and $P$ denotes their order. 

The integral $I_{m_2\cdots m_{4n}}$ can be easily calculated at leading order
\begin{eqnarray}
&& I_{m_2\cdots m_{4n}}:= \prod \limits_{h=2}^{2n} \int \sqrt{-g}\,  d^4 x_{h} \prod \limits_{s=1}^{2n}  S_{m_{2s}}(x_{s}, x_{|s+1|_4}) \nonumber \\
 && =\frac{1}{4^{2n} \pi^{4n}\, a^3 } \prod \limits_{h=2}^{2n} \int  d^4 x_{h} \prod \limits_{s=1}^{2n}  \partial_{m_{2s}}^{(s)} \frac{1}{\Delta_{++}^2(x_{s}, x_{|s+1|_4})}\nonumber\\
 &&=\frac{1}{a^3\, (4 \pi^2)^{2n}\, \mu^{2(n-2)} (2n-1)!  }\,.
\end{eqnarray}

\end{document}